\documentclass[prb,twocolumn,showpacs,amsmath,amssymb]{revtex4}

\usepackage{graphicx}
\usepackage{dcolumn}
\usepackage{bm}

\begin{document}

\title{Phase Diagrams of the Two-Orbital Hubbard Model with Different Bandwidths}

\author{Kensuke Inaba}
\author{Akihisa Koga}%

\affiliation{%
Department of Applied Physics, Osaka University, Suita, Osaka 565-0871, Japan
}%

\date{\today}

\begin{abstract}
We investigate the two-orbital Hubbard model in infinite dimensions
by means of the self-energy functional method.
By calculating the entropy, susceptibility, and quasi-particle weight
at zero temperature, we determine the phase diagram for the system
with same and different bandwidths,
which is compared with that obtained recently.
It is clarified that orbital fluctuations 
play a key role in controlling 
the nature of the Mott transitions in the system.
\end{abstract}


\preprint{APS/123-QED}

\maketitle

\section{Introduction}

Transition metal oxides have attracted much interest in their various aspects.
\cite{ImadaRev,TokuraScience}
Among them, heavy fermion behavior in $d$-electron systems
stimulates experimental and theoretical investigations.
One of typical examples is the lithium vanadate $\rm LiV_2O_4$,\cite{Kondo97} 
where heavy fermion behavior with
the enhanced specific-heat coefficient $\gamma_e=0.42J/mol K^2$ 
has been observed at low temperatures.
It has been suggested that the large mass enhancement
 in this system may originate from
geometrical frustration.\cite{Kaps01,Isoda00,Fujimoto01}
Recently, it has been pointed out that degenerate 
orbitals in the $t_{2g}$ subshell 
also play an important role in realizing heavy fermion behavior in 
$\rm LiV_2O_4$.\cite{Tsunetsugu02,Yamashita03}
Another example is the isovalent ruthenate arroy $\rm Ca_{2-x}Sr_xRuO_4$.
In the compound,
the substitution of $\rm Ca^{2+}$ ions for $x>0.5$
realizes heavy fermions in the $t_{2g}$ subshell, \cite{Nakatsuji} 
where
some of physical quantities have a cusp singularity around $x=0.5$.
Furthermore, the unexpected $s=1/2$ moment 
per $\rm Ru$-ion coexisting with the metallic state has been
observed in that region,  \cite{Nakatsuji} 
which suggests the existence of the orbital-selective Mott transition (OSMT).
\cite{Anisimov02,Fang04}
This type of the Mott transition has also been suggested to occur in the compound 
$\rm La_{n+1}Ni_nO_{3n+1}$,\cite{LaNiO,Kobayashi96}
stimulating further experimental investigations on the multiorbital systems.

In the above compounds, electron correlations in the system 
with orbital degeneracy are important in understanding  
heavy fermion behavior.
In particular, the concept of the OSMT provides a new paradigm of
metal-insulator transitions in the multiorbital systems.
Although intensive theoretical studies on these topics have recently been done,
\cite{Kim90,Kotliar96,Rozenberg97,Bunemann:Gutzwiller,Hasegawa98,Held98,Han98,Momoi98,Klejnberg98,Imai01,Koga:ED_DMFT,Florens02,Florens02b,Oudovenko02,Ono03,Tomio05,Pruschke05,Sakai04,Liebsch03,Liebsch03L,Liebsch04,Sigrist04,Fang04,Anisimov02,Koga04,Koga05,Koga05Pr,KogaRev,Ferrero05,Medici05,Arita05,Knecht05,Liebsch05,Biermann05,Inaba05,Inaba05os,Ruegg05,Song05}
there are some controversial conclusions
even in the simplest multiorbital model.
In fact, in the infinite-dimensional two-band system with equivalent orbitals,
the existence of the Hund coupling induces 
the first-order Mott transition to the insulating phase.
\cite{Bunemann:Gutzwiller,Ono03,Inaba05}
However, recent results obtained by 
dynamical mean-field theory (DMFT)\cite{GeorgesRev,KotliarPT}
combined with the numerical renormalization group 
claim that the Mott transition is of second-order 
in a certain parameter region.\cite{Pruschke05}
In addition,
there are some open questions 
in the system with different bandwidths.
It has been pointed out that the double OSMTs occur in general,
which merge to a single Mott transition only under the special condition.
\cite{Koga04,Liebsch03,Liebsch03L,Liebsch04}
On the other hand, it has been proposed that the double transitions always occur
if the system has a large difference of the bandwidths.\cite{Medici05,Ferrero05}
Therefore, it is highly desired to discuss the nature of the Mott transition 
in order to clarify what triggers the single or double transitions 
in the system with different bandwidths.

Motivated by this, we investigate the Mott transitions 
in the multiorbital systems.
In particular, we focus on orbital fluctuations in this paper,
which may play an important role in stabilizing the metallic state in the system.
\cite{Koga:ED_DMFT,Koga05}
By making use of the self-energy functional approach (SFA)
proposed by Potthoff,\cite{Potthoff03a,Potthoff03b}
we confirm some previous works \cite{Medici05,Ferrero05} and 
determine the phase diagrams.
We examine how orbital fluctuations are affected by 
the Hund coupling and/or the difference of the bandwidths 
to clarify that the enhanced orbital fluctuations 
make single Mott transition stable
against double OSMTs.

This paper is organized as follows.
In Sec. \ref{sec:model}, we introduce the two-orbital Hubbard model 
and  briefly summarize the SFA.\cite{Potthoff03a,Potthoff03b}
The nature of the Mott transitions in the two-orbital system 
with same bandwidths is discussed in Sec. \ref{sec:same_bandwidths}.
We also discuss how the OSMT is realized in the system 
with different bandwidths in Sec. \ref{sec:different_bandwidths}.
A brief summary is given in Sec. \ref{sec:summary}.

\section{Model and method}\label{sec:model}
We consider the two-orbital Hubbard model with different bandwidths,
which is given by the following
Hamiltonian, ${\cal H}={\cal H}_0+{\cal H}^\prime$,
${\cal H}'=\sum_i {\cal H}_i^\prime$ with
\begin{eqnarray}
{\cal H}_0&=&\sum_{<i,j>, \alpha, \sigma}
\left( t_\alpha -\mu\delta_{ij}\right)
          c^\dag_{i\alpha\sigma} c_{j\alpha\sigma},\\
{\cal H}_i^\prime&=& \!\!\!\!
                  U \sum_{\alpha}
                n_{i \alpha \uparrow} n_{i \alpha \downarrow}
                  +\sum_{\sigma \sigma^\prime}
                 (U^\prime - \delta_{\sigma\sigma^\prime}J)
                n_{i 1 \sigma} n_{i 2 \sigma^\prime}
                \nonumber\\
          &-&  \!\!\!\!\! J  (c^\dag_{i 1 \uparrow}c_{i 1 \downarrow}
                c^\dag_{i 2 \downarrow}c_{i 2 \uparrow}
                +c^\dag_{i 1 \uparrow}c^\dag_{i 1 \downarrow}
                c_{i 2 \uparrow}c_{i 2 \downarrow}+H.c.)
          \label{eq:model},
\end{eqnarray}
where $c^\dag_{i\alpha\sigma}(c_{i\alpha\sigma})$ creates (annihilates) 
an electron with spin $\sigma (=\uparrow, \downarrow)$ and orbital
$\alpha (=1,2)$ at the $i$ th site, 
and $n_{i\alpha\sigma}$ is the number operator.
Here, $t_\alpha$ denotes the hopping integral for the $\alpha$th orbital, 
$\mu$ the chemical potential, 
$U (U')$ the intra-orbital (inter-orbital) Coulomb interaction, 
and $J$ the Hund coupling including the spin-flip and pair-hopping terms. 
In the paper, we impose the condition $U=U'+2J$ due to
the rotational symmetry of degenerate orbitals.


To discuss the Mott transitions 
in the multiorbital systems,
we make use of the SFA.
Since this method is based on the variational principle,
it has an advantage to discuss the properties of the Mott transition systematically.
In fact, it has successfully been applied to various systems 
such as Hubbard models in infinite dimensions
\cite{Potthoff03a,Potthoff03b,Pozgajcic04,Balzer05,Koller05} 
and finite dimensions.
\cite{Potthoff03L,Aichhorn04,Dahnken04,Tong05,Senechal05,Aichhorn05,Aichhorn05EL}

In this approach, the ground potential $\Omega$ is given as,
\begin{eqnarray}
\Omega [{\boldsymbol \Sigma}]&=&F[{\boldsymbol \Sigma}] + 
{\rm Tr}\ln [-({\bf G}_0^{-1}-{\boldsymbol \Sigma})^{-1}]\label{eq:LW},
\end{eqnarray}
where $F[{\boldsymbol \Sigma}]$ is the Legendre transformation of 
the Luttinger-Ward potential. \cite{Luttinger}
${\bf G}_0$ and ${\boldsymbol \Sigma}$ are the 
bare Green function  and the self-energy, respectively.
Here the Dyson equation ${\bf G}^{-1}={\bf G}_0^{-1} -{\boldsymbol \Sigma}$
is obtained under the condition 
${ \partial \Omega[{\boldsymbol \Sigma}] }/{ \partial {\boldsymbol \Sigma}} =0$,
\cite{Luttinger}
where ${\bf G}$ is the full Green function. 
We wish to note that
the potential $F[{\boldsymbol \Sigma}]$ does not depend on the detail of 
the non-interacting Hamiltonian.\cite{Potthoff03a}
This allows us to introduce a reference system 
with the same interacting term.
The Hamiltonian is explicitly given by
${\cal H}_{\rm ref}({\bf t}')={\cal H}_0({\bf t}')+{\cal H}'$ 
with the parameter matrix ${\bf t}'$.
Then we obtain the grand potential as,
\begin{eqnarray}
\Omega [{\boldsymbol \Sigma}({\bf t}^\prime)]&=&\Omega({\bf t^\prime})
\nonumber\\
    &+&{\rm Tr}\ln
    \left[-(\omega+\mu-{\bf t}-{\boldsymbol \Sigma}
({\bf t^\prime}))^{-1}\right]
    \nonumber\\
    &-&{\rm Tr}\ln
    \left[-(\omega+\mu-{\bf t}'-{\boldsymbol \Sigma}
({\bf t^\prime}))^{-1}\right],\label{eq:omega_SFA}
\end{eqnarray}
where $\Omega({\bf t}')$ and ${\boldsymbol \Sigma}({\bf t}')$ are 
the grand potential and the self-energy for the reference system.
The condition 
$\partial \Omega[{\boldsymbol \Sigma({\bf t}')}]/\partial {\bf t}'=0$
gives us an appropriate reference system 
${\cal H}_{\rm ref}({\bf t}')$ 
in the framework of the SFA.

If the SFA is applied to infinite-dimensional correlated electron systems,
an Anderson impurity model is one of the most appropriate 
reference systems,
which is given by
\begin{eqnarray}
{\cal H}_{\rm ref}&=&\sum_i{\cal H}_{\rm ref}^{(i)},\\
  {\cal H}_{\rm ref}^{(i)}&=&\sum_{\alpha \sigma } \varepsilon^{(i)}_{0\alpha}
          c^\dag_{i\alpha\sigma} c_{i\alpha\sigma}+\sum_{k=1}^{N_b}\sum_{\alpha \sigma} 
          \varepsilon^{(i)}_{k \alpha}
          a^{(i)\dag}_{k\alpha\sigma}a^{(i)}_{k\alpha\sigma}\nonumber\\
          &+&\sum_{k=1}^{N_b}\sum_{\alpha \sigma}V^{(i)}_{k \alpha}
          (c^\dag_{i\alpha\sigma}a^{(i)}_{k\alpha\sigma}+H.c.)+{\cal H}_i',\label{eq:ref_model}
\end{eqnarray}
where $a^{(i)\dag}_{k\alpha\sigma}(a^{(i)}_{k\alpha\sigma})$ 
creates (annihilates) an electron with spin $\sigma$ and orbital $\alpha$ 
at the $k(=1,2,\cdots N_b)$th site, 
which is connected to the $i$th site in the original lattice. 
In the limit of $N_b\to \infty$,
the condition $\partial \Omega[{\boldsymbol \Sigma({\bf t}')}]/\partial {\bf t}'=0$
reproduces self-consistent equations of DMFT.
\cite{Potthoff03a,Potthoff03b,Pozgajcic04}

The grand potential per site is rewritten as,
\begin{eqnarray}
\Omega /L&=&\Omega_{\rm imp}
    -2 \sum _{\alpha}\sum_m F( \omega'_{\alpha m}) -2\sum_\alpha\sum_{k=1}^{N_b} F(\omega^b_{k\alpha}) \nonumber\\
    &+&2 \sum _{\alpha}\sum_m  \int^{-\infty}_{\infty}dz 
                             \rho_\alpha(z) F[ \omega_{\alpha m}(z)],\label{eq:omg}\\
 F(x)&=&-T \ln [1+\exp(-x/T)]
\end{eqnarray}
where $\Omega_{\rm imp}$ is the grand potential for the reference system.
$\omega_{\alpha m}(z)$ $[\omega'_{\alpha m}]$
is the pole of Green function $G$ $(G')$ for
the original (reference) system and
$\omega^b_{k\alpha}=\varepsilon_{k\alpha}-\mu$. 
The Green function of the reference system is given as,
\begin{eqnarray}
G'_{\alpha}(\omega)&=&[\omega+\mu-\varepsilon_{0\alpha}
           -\Delta_\alpha(\omega)-\Sigma_{\alpha}(\omega)]^{-1}, \\
\Delta_\alpha(\omega)&=&\sum_{k=1}^{N_b} 
\frac{V_{k\alpha}^2}{\omega-\omega^b_{k\alpha}} 
\end{eqnarray}
where $\Sigma_\alpha(\omega)$ is the self-energy for the $\alpha$th orbital. 
On the other hand, the full Green function is given as,
\begin{equation}
G_\alpha(\omega;z)=[\omega+\mu-z-\Sigma_\alpha(\omega)]^{-1}.
\end{equation}
Note that the following differential equation is efficient to deduce 
the poles $\omega_{\alpha m}(z)$ in the Green function,
\begin{eqnarray}
\frac{d\omega_{\alpha m}(z)}{dz}=\left(1-\frac{\partial\Sigma_\alpha(\omega)}{
\partial \omega}\right)^{-1}_{\omega=\omega_{\alpha m}(z)}.
\end{eqnarray}
By solving these equations, we estimate the grand potential numerically 
to discuss the effect of electron correlations.

%

To clarify the nature of Mott transitions, 
we calculate various physical quantities.
In the metallic phase, the quasi-particle weight for the $\alpha$th orbital, 
$Z_\alpha$, 
is useful to discuss how the Fermi liquid states are renormalized by 
the Coulomb interactions.
This quantity 
is proportional to the inverse of the effective mass,
which is defined as,
\begin{eqnarray}
Z^{-1}_\alpha&=&\left(1-\frac{\partial \Sigma_\alpha(\omega)}{\partial \omega}\right)_{\omega=0}.
\end{eqnarray}
Furthermore, we calculate the local susceptibilities, which may clarify
how spin, orbital and charge fluctuations affect 
the stability of the metallic state.
These read
\begin{eqnarray}
\chi_s&=&\int^\beta_0 \langle {\cal T} [n_\uparrow(\tau)-n_\downarrow(\tau)][n_\uparrow(0)-n_\downarrow(0)]\rangle d\tau,\nonumber\\
\chi_o&=&\int^\beta_0 \langle {\cal T} [n_1(\tau)-n_2(\tau)][n_1(0)-n_2(0)]\rangle d\tau,\nonumber\\
\chi_c&=&\int^\beta_0 \langle {\cal T} [n(\tau)-2][n(0)-2]\rangle d\tau,\label{eq:susceptibilities}
\end{eqnarray}
where $n=\sum_{\alpha\sigma} n_{\alpha\sigma}$, 
$n_{\sigma(\alpha)}=\sum_{\alpha(\sigma)} n_{\alpha\sigma}$ and 
$n_{\alpha\sigma}=c^\dag_{i\alpha\sigma}c_{i\alpha\sigma}$, 
${\cal T}$ the time ordered operator, 
$A(\tau)=e^{-{\cal H}\tau}Ae^{{\cal H}\tau}$ and 
$\beta$ the inverse temperature.
Note that in infinite dimensions, 
these local quantities coincide with those for the reference system 
with $N_b \to \infty$.
\cite{GeorgesRev}
In the paper, calculating these quantities approximately in terms of 
the reference system with a finite $N_b$,
we discuss how fluctuations affect the stability of the metallic state.
We also calculate the residual entropy in the system 
to characterize the Mott insulating phase,
where localized electrons are realized
with the free spins. 
It is given as,
\begin{eqnarray}
S&=&-\frac{d\Omega}{dT}\\
&=&S_{\rm imp}
    +2 \sum _{\alpha}\sum_m F'( \omega'_{\alpha m}) 
      +2\sum _{\alpha}\sum_{k=1}^{N_b} F'( \omega^b_{k\alpha }) \nonumber\\
    &-&2 \sum _{\alpha}\sum_m  \int^{-\infty}_{\infty}dz 
                             \rho_\alpha(z) F'[ \omega_{\alpha m}(z)],\label{eq:entropy}\\
 F'(x)&=&[F(x)-(x-\partial x/\partial T)f(x)]/T,\\
 f(x)&=&1/(1+\exp(x/T)),
\end{eqnarray}
where $S_{\rm imp}=-\partial \Omega_{\rm imp}/\partial T$ is 
the entropy of the reference system and
we have used the condition $\partial \Omega / \partial {\bf t}'=0$.
Note that the temperature derivative of the poles 
for the reference system is zero:
$\partial \omega'_{\alpha m}/\partial T=0$.
By estimating $\partial \omega_{\alpha m}(z) / \partial T$
carefully,\cite{Pozgajcic04} we can calculate the entropy for the system.

Here, we use the semicircular density of states 
$\rho_\alpha(\omega)=4/\pi W_\alpha \sqrt{1-(2x/W_\alpha)^2}$,
where $W_\alpha(=4t_\alpha)$ is the bandwidth for the $\alpha$th orbital,
which corresponds to an infinite coordination Bethe lattice.
In this paper, we restrict our discussions to the paramagnetic case 
in the half-filled system, by setting the chemical potential $\mu=U/2+U'-J/2$.
In the following, by varying the ratio of the bandwidths $R\equiv W_1/W_2 (<1)$
with a fixed $t_2=1$ (that is an energy unit), 
we proceed to discuss the Mott transitions 
in the two-orbital model.


\section{Mott transition in the system with same bandwidths}\label{sec:same_bandwidths}

Let us consider the two-orbital system with same bandwidths.
Mott transitions in the system have been discussed so far, by combining DMFT
with numerical techniques.\cite{Kim90,Kotliar96,Rozenberg97,Bunemann:Gutzwiller,Hasegawa98,Held98,Han98,Momoi98,Klejnberg98,Imai01,Koga:ED_DMFT,Florens02,Florens02b,Oudovenko02,Ono03,Tomio05,Pruschke05,Sakai04,Liebsch03,Liebsch03L,Liebsch04,Sigrist04,Fang04,Anisimov02,Koga04,Koga05,Koga05Pr,KogaRev,Ferrero05,Medici05,Arita05,Knecht05,Liebsch05,Biermann05,Inaba05,Inaba05os,Ruegg05,Song05}
An exact diagonalization study claimed that the first-order transition occurs 
with the jump in physical quantities.\cite{Ono03}
In contrast, it was suggested that the second-order transition occurs 
in a certain parameter region by means of the numerical renormalization group.
\cite{Pruschke05}
In this section, to resolve
the controversial conclusions for the properties of the Mott transition,
we make use of the SFA with $N_b=1$.
At the end of the section,
we check the validity of our analysis by comparing 
the results of $N_b=1$ with those of $N_b=3$.

\begin{figure}[htb]
\includegraphics[width=\linewidth]{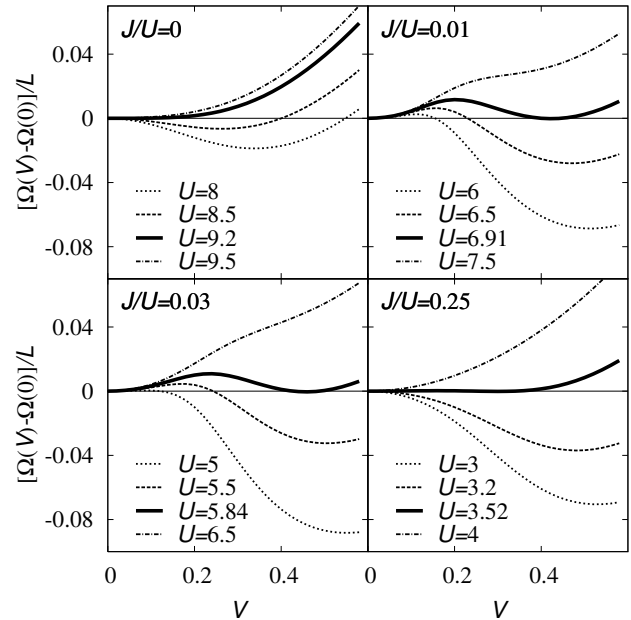}
\caption{The grand potential as a function of the variational parameter $V$.
Thick lines correspond to the critical value $U_c$.
}\label{fig:omega}
\end{figure}

We first calculate the ground potential $\Omega$ 
as a function of $V\equiv V_{1,1}=V_{1,2}$ at zero temperature,
as shown in Fig. \ref{fig:omega}.

When $U$ is small, it is found that 
the grand potential has a minimum at finite $V$.
The hybridization $V$ in the reference system, 
roughly speaking, represents the effective bandwidth 
for the original system. 
Therefore, 
the metallic state is stabilized in this case.
It is seen that the increase of the Coulomb interaction 
with a fixed ratio $J/U=0$ shifts 
the stationary point toward the origin continuously.
This implies that the effective bandwidth is gradually decreased, and 
the transition occurs to the Mott insulating phase at $U=U_c$.
In contrast, the introduction of the Hund coupling 
leads to different behavior.
For instance, we focus on the system with the weak Hund coupling $J/U=0.03$.
When $U=5$, a minimum appears at $V\sim0.58$,
where the metallic ground state is realized.
The increase of the Coulomb interaction induces
another minimum around $V\sim0$, which represents the insulating state.
This double-well structure in the grand potential
suggests the existence of the first-order Mott transition in the system.
By comparing the energy for each state, 
we determine the critical value $U_c=5.84$.
A similar structure is also observed in the case $J/U=0.25$. However,
the potential barrier between these two states becomes small
at the first-order transition point, as shown in Fig. \ref{fig:omega}.
Therefore, it is expected that the singularity 
characteristic of the first-order transition
is more difficult to be observed in this case.

\begin{figure}[htb]
\includegraphics[width=\linewidth]{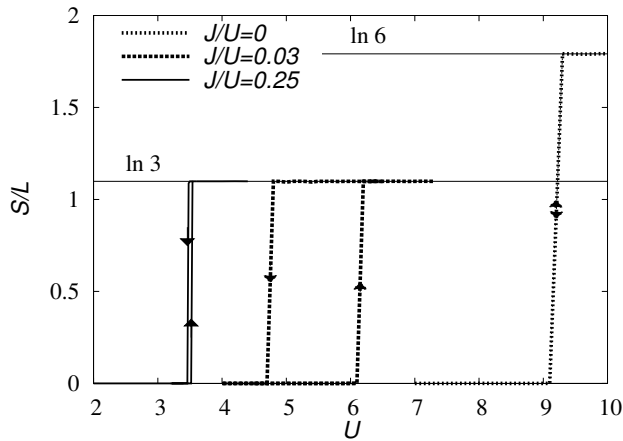}
\caption{The entropy per site as a function of $U$.
}\label{fig:entropy}
\end{figure}

These features mentioned above can be seen in the physical quantities.
Here, we show the entropy per site $S/L$ in Fig. \ref{fig:entropy}.
When $J/U=0$, 
the increase of the Coulomb interaction triggers the Mott transition
at $U=U_c$, where the entropy jumps up to $S/L=\ln 6$.
The value is explained by the fact that
sixth degenerate states are lowest in the atomic limit.
On the other hand, the Hund coupling forms the triply degenerate states 
in the limit.
Therefore, when $J/U\neq 0$, $S/L=\ln 3$ appears in the insulating phase.
It is found that the second-order transition occurs in the system $J/U=0$, 
while the first-order transition with the hysteresis
occurs in the system with the Hund coupling.
For example, when $J/U=0.03$, 
the metallic state is stabilized up to the critical point $U_c^M=6.12$, 
while the insulating state is stabilized down to the critical point $U_c^I=4.82$.
Note that these critical points $U_c^M$ and $U_c^I$ 
induced by the Hund coupling at zero temperature are different from 
critical points $U_{c1}$ and $U_{c2}$,
which are defined as Mott critical points at finite temperatures.
Therefore, an inequality $U_c^I \le U_c^M$ is not necessarily satisfied,
which will be discussed in detail in the next section.

\begin{figure}[htb]
\includegraphics[width=\linewidth]{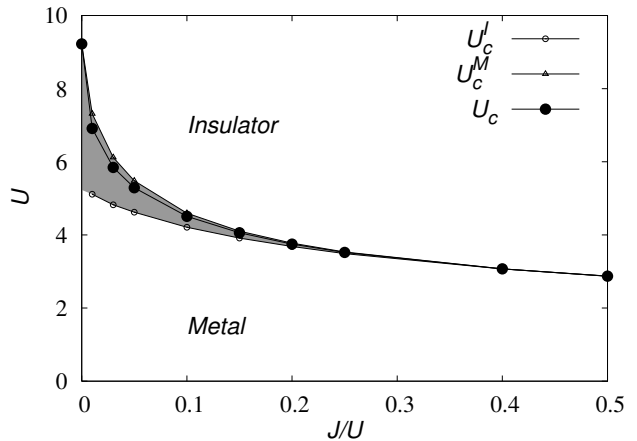}
\caption{The zero-temperature phase diagram for 
the degenerate Hubbard model with same bandwidths $R=1$. 
$U_c^I$ ($U_c^M$) is the critical point where
the insulating (metallic) state disappears.
$U_c$ represents the first order transition points and
the filled region represents the coexisting state.
}\label{fig:Up_U_W10}
\end{figure}

By performing similar calculations, we obtain the phase diagram
as shown in Fig. \ref{fig:Up_U_W10}.
We find that the phase boundary $U_c^M$ is always larger than
$U_c^I$ except for the special condition $J/U=0$.
This implies that the first-order Mott transition, in general, 
occurs in the two-orbital system with same bandwidths.
When the system has the special condition $J/U=0$,
the critical point $U_c^I(=U_c^M=9.24)$ is different from 
the extrapolation of the curve $U_c^I$.\cite{Inaba05}
This exception may originate from 
the discontinuity of the residual entropy for the Mott insulating phase.
It is also found that the coexisting region bounded by these two lines
shrinks with the increase of the Hund coupling $J$.
This interesting feature may be due to orbital fluctuations,
which sometimes play an important role in understanding the Mott transitions
in the two-orbital system.

\begin{figure}[htb]
\includegraphics[width=\linewidth]{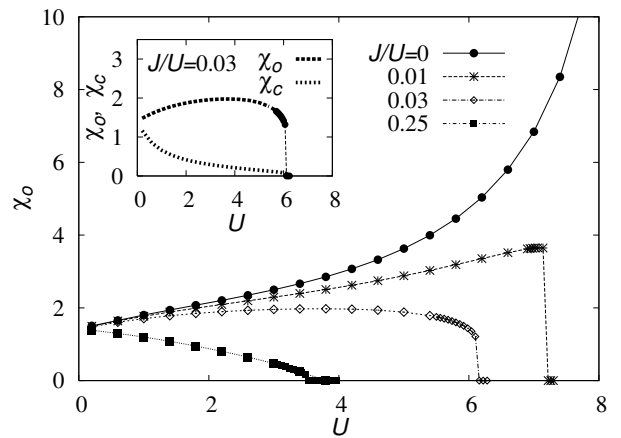}
\caption{The local orbital susceptibility in the system with same bandwidths.
The inset shows orbital and charge susceptibilities in the case $J=0.03U$.}
\label{fig:sus_R1}
\end{figure}

To make this point clear, we also calculate the local orbital susceptibility
in the metallic state $(U<U_c^M)$, 
as shown in Fig. \ref{fig:sus_R1}.
When $J/U=0$, the increase of the Coulomb interaction enhances
the spin and orbital susceptibilities 
(the spin susceptibility is not shown).
Eventually, these quantities diverge simultaneously at the transition point.
On the other hand, when one turns on the Hund coupling $(J/U=0.01)$,
different behavior appears.
The introduction of the Coulomb interaction first enhances orbital fluctuations,
by reflecting the high symmetry $U\sim U'$.
Further increase of the interactions renormalizes electrons in both orbitals,
where the Hund coupling has a tendency to suppress
orbital fluctuations.
In fact, in contrast to the case $U=U'$,
the orbital susceptibility is much smaller around the Mott critical point.
We also find the sudden decrease of the orbital susceptibility 
where the first-order Mott transition occurs.
When the system is located far from the condition $U=U'$, 
the increase of the Coulomb interaction suppresses orbital fluctuations 
strongly. 
Therefore, a tiny jump appears in the curve at $U_c^M=3.54$ $(J/U=0.25)$.

The obtained results may shed light on the nature of the Mott transitions in
the two-orbital systems with same bandwidths.
When $J/U=0$, the second-order Mott transition occurs to the insulating
phase.
On the other hand, the introduction of the Hund coupling induces 
the first-order transition with the hysteresis in the physical quantities.
It is also found that as $J$ is increased, orbital fluctuations are suppressed
strongly, where the singularity for the transition becomes obscure.
In this case, the system can be regarded as the system with 
independent orbitals.
Therefore, the OSMT is expected to occur if the bandwidths are 
different from each other, which will be discussed in the next section.
We wish to comment on the related recent work.\cite{Ono03,Pruschke05}
The introduction of the Hund coupling is known to induce the first-order
transition, while the nature of transitions in the large $J$ region
has not been clarified up to now.
In the region, the first- (second-)order transition was suggested 
by the exact diagonalization (numerical renormalization group).
On the other hand, our systematic analysis shows the jump singularity 
in the curve for each physical quantity, except for the $J=0$ case.
Therefore, we believe that further increase of the Hund coupling 
does not induce the second-order Mott transition in the system with same bandwidths.

\begin{figure}[htb]
\includegraphics[width=\linewidth]{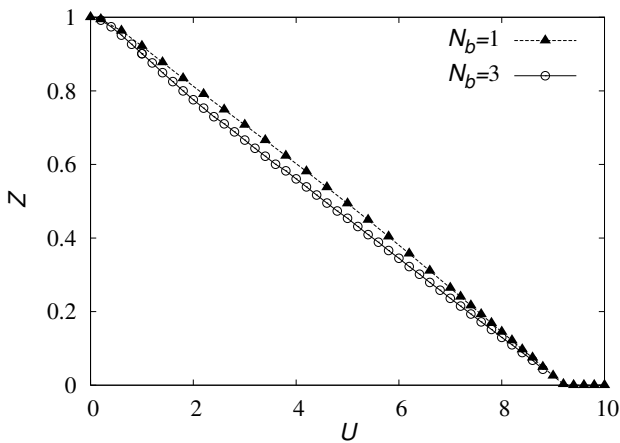}
\caption{The quasiparticle weight $Z$ as a function of $U$ 
obtained from the reference systems $N_b=1$ and $N_b=3$ when $J/U=0$.
}\label{fig:rnm_FL}
\end{figure}

Before closing this section, we would like to check the validity of our analysis.
In Fig. \ref{fig:rnm_FL}, we show the quasiparticle weights 
obtained by the SFA with $N_b=1$ and $N_b=3$.
If the Coulomb interactions are introduced, the quasi-particle weight
is decreased from unity.
In this region, it is found that the quasi-particle weight 
obtained for the system $N_b=1$ is
slightly larger than that for the system $N_b=3$.
On the other hand, when the system approaches the Mott transition point,
the quasi-particle weight is hardly affected by the number of 
the sites for the reference system.
This result may be explained by the fact that
the simplified reference system ($N_b=1$) can describe 
the low-energy physics around the Fermi level and thereby 
we can determine the critical point quantitatively.
\cite{Potthoff03a,Potthoff03b,Pozgajcic04} 
However, the reference system $N_b=1$ 
does not describe the high energy part  properly, resulting in 
the overestimate of $Z$ in the intermediate region.

\begin{figure}[htb]
\includegraphics[width=\linewidth]{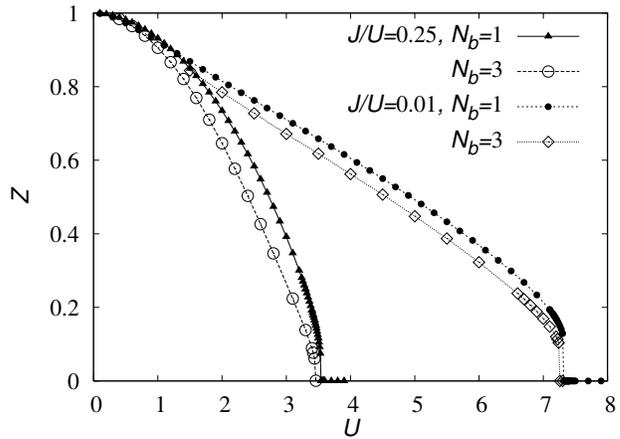}
\caption{The quasiparticle weight $Z$ as a function of $U$ in the cases 
$J/U=0.01$ and $J/U=0.25$. }\label{fig:rnm_FL_J}
\end{figure}

Similar behavior is also observed in the finite $J$ case, 
as shown in Fig. \ref{fig:rnm_FL_J}.
As the Hund coupling is increased, 
the Fermi liquid state is stabilized up to 
$U_c^M\sim 7.3$ and $3.5$ at $J/U=0.01$ and $0.25$ respectively, 
where the first-order transition occurs.
It is found that the jump singularity for $N_b=3$
is weaker than that for $N_b=1$, which may expect that the transition becomes 
of second order in the $N_b \to \infty$ case. 
According to the results for the single-band model,
\cite{Potthoff03a,Potthoff03b,Pozgajcic04}
however, the physical quantities for $N_b=3$ 
nearly equal those for $N_b > 3$.
Therefore, we believe that the nature of the Mott transition 
is hardly affected by $N_b$, and
the first-order transition remains in the $N_b \to \infty$ case.
In addition,
we also find that 
the critical point $U_c^M$ weakly 
depends on the number of sites $N_b$, implying 
that the obtained phase boundaries shown in Fig. \ref{fig:Up_U_W10}
have been determined rather well even by the reference system $N_b=1$.

\section{Orbital-selective Mott transitions vs. Single Mott transition}
\label{sec:different_bandwidths}

In this section, we consider Mott transitions in the two-orbital system 
with different bandwidths $(R\neq1)$. 
\cite{Liebsch03,Liebsch03L,Liebsch04,Sigrist04,Fang04,Anisimov02,Koga04,Koga05,Koga05Pr,KogaRev,Ferrero05,Medici05,Arita05,Knecht05,Liebsch05,Inaba05os,Biermann05,Ruegg05,Tomio05,Song05}
It has been suggested that in the system with $R=0.5$, 
the OSMT occurs in the case $J/U=0.25$, while
a single Mott transition occurs in the case $J/U=0$.\cite{Koga04,Liebsch03L}
On the other hand,
it was recently claimed that the double transitions always occur in the small $R$ case.
\cite{Medici05,Ferrero05}
By means of the SFA with $N_b=1$,
we determine the detailed phase diagram 
to discuss the nature of the Mott transitions systematically.


First we consider the system without Hund coupling ($J/U=0$).
In the system with same bandwidths ($R=1$),
the second-order Mott transition occurs at $U_c=9.24$,
where the quasi-particle weight continuously reaches zero and 
the orbital susceptibility diverges.
\begin{figure}[htb]
\includegraphics[width=\linewidth]{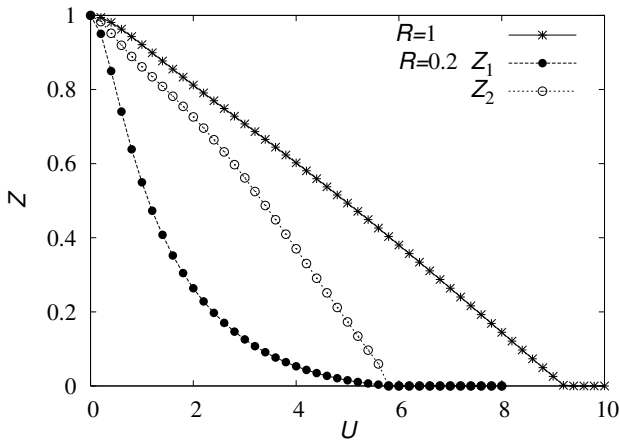}
\caption{The quasi-particle weights for $\alpha$ th orbital $Z_\alpha$ when $R=1$ and $0.2$.
}\label{fig:rnm_J0}
\end{figure}
\begin{figure}[htb]
\includegraphics[width=\linewidth]{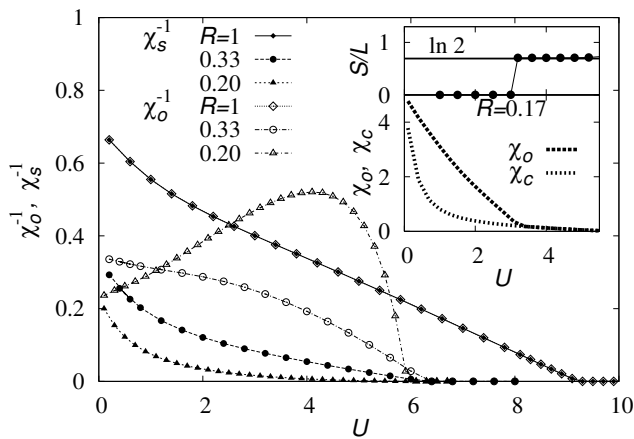}
\caption{The inverse of orbital and spin local-susceptibilities
as a function of $U$ when $J/U=0$.
The inset shows orbital and charge susceptibilities, and the entropy per site
when $R=0.17$.
}\label{fig:sus_J0}
\end{figure}
Away from the condition ({\it e.g.} $R=0.2$),
the introduction of the interaction decreases the quasi-particle weights 
$Z_1$ and $Z_2$
in  different ways reflecting the difference of bandwidths,
as shown in Fig \ref{fig:rnm_J0}. 
This enhances the spin fluctuations and suppresses
orbital fluctuations.
However, when the system approaches the Mott transition point,
the effect of the bare bandwidths is diminished 
due to the strong renormalization, 
which enhances both spin and orbital fluctuations.
Therefore, in the case $R=0.2$, 
monotonic (non-monotonic) behavior appears 
in the curves of the spin (orbital) susceptibility.
As a consequence, a single Mott transition occurs, 
where the quasi-particle weight for each orbital vanishes simultaneously.\cite{Koga04}
On the other hand,  
somewhat different behavior appears in the case $R<R_c(=0.192)$, 
as shown in the inset of Fig. \ref{fig:sus_J0}.
If the Coulomb interactions are increased,
orbital fluctuations are strongly suppressed
due to the difference of the effective Coulomb interactions. 
In this case, the orbital susceptibility never diverges, 
which suggests the existence of the OSMT, 
as discussed by Medici {\it et al}.\cite{Medici05} and Ferrero {\it et al}.\cite{Ferrero05}.
In fact, the OSMT yields a localized spin $s=1/2$ at each site.
Therefore,
the residual entropy $S/L=\ln 2$ appears and 
the orbital susceptibility merges to the charge susceptibility at $U_c^M\sim 3.2$,
as shown in the inset of Fig. \ref{fig:sus_J0}.
Further increase of the Coulomb interaction decreases the charge and orbital 
susceptibilities and another Mott transition occurs at $U_c^I\sim 5.9$.

\begin{figure}[htb]
\includegraphics[width=\linewidth]{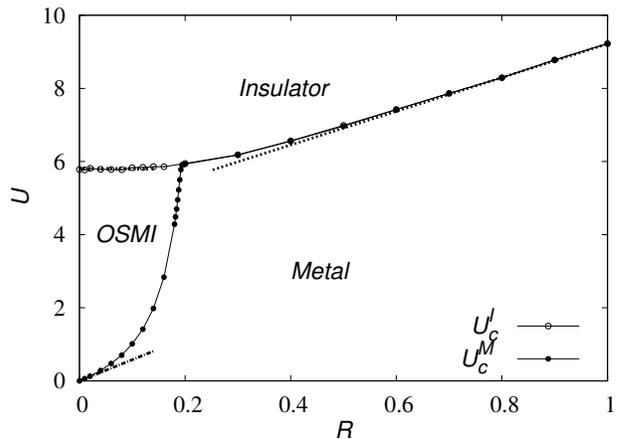}
\caption{The phase diagram for the case $J/U=0$. 
$U_c^I$ ($U_c^M$) is the critical point where
the  insulating (metallic) state becomes unstable
with increasing (decreasing) $U$.
The dotted lines denote the phase boundaries obtained by 
the simple estimation (see text).
}
\label{fig:R_U_J0}
\end{figure}

By estimating the critical points systematically,
we obtain the phase diagram shown in Fig. \ref{fig:R_U_J0}.
When $R=1$, the model is reduced to the system with same bandwidths, where
the Mott transition occurs between the metallic and Mott insulating phases.
Even when the system has two distinct orbitals,
the single Mott transition  still occurs ($U_c^M=U_c^I$)
owing to enhanced orbital fluctuations
within the region $(R_c<R<1)$.
In this region, the phase boundary is simply deduced 
by the Gutzwiller approximation as, 
$U_{c} \propto \int d\epsilon \epsilon \rho(\epsilon) = (1+R)/2$,
which is in good agreement with the results obtained by the SFA.
On the other hand, in the region $R\lesssim R^c$, 
the double Mott transitions occur ($U_c^M<U_c^I$) when the interaction is varied.
Then the orbital-selective Mott insulating (OSMI) phase appears
between the metallic and insulating phases.\cite{Medici05,Ferrero05}
In the case, the system may be regarded as two independent 
single-band systems.
In fact, the phase boundaries $U_c^I \sim u_{c} R$ and $U_c^M \sim u_{c}$,
where $u_{c}=5.85$ is the critical point for the single-band model,
\cite{GeorgesRev,Potthoff03b}
are in good agreement with our SFA results 
in the case $R \ll 1$, as shown in Fig. \ref{fig:R_U_J0}.
The phase diagram is consistent with that obtained
by Medici {\it et al}.\cite{Medici05} and Ferrero {\it et al}.\cite{Ferrero05}.



Next, we consider the effect of the Hund coupling in the system 
with different bandwidths $R=0.5$.
\begin{figure}[htb]
\includegraphics[width=\linewidth]{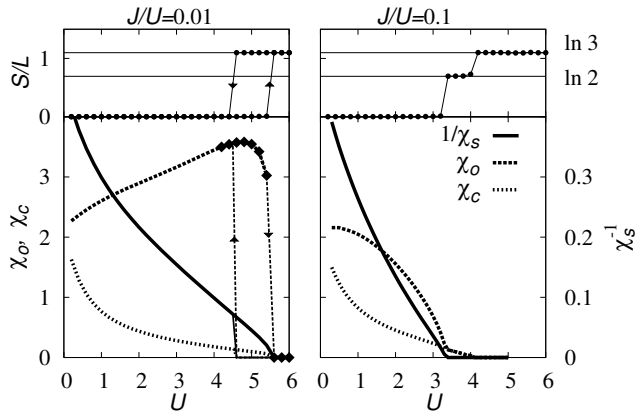}
\caption{The local susceptibilities and the entropy per site
in the system with different bandwidths $R=0.5$.}\label{fig:sus_R2}
\end{figure}
In Fig. \ref{fig:sus_R2}, we show the physical quantities as a function of 
$U$ with a fixed ratio $J/U$.
When the Hund coupling is small $J/U=0.01$, the interaction 
renormalizes electrons
depending on each bandwidth,
and it triggers the first-order single transition 
to the Mott insulating phase at $U_c^M=5.47$.
Then, the singularity appears in the curve of the susceptibility and
the entropy.
This implies that the existence of the Hund coupling induces 
the first-order Mott transition with the hysteresis, 
where the Mott insulating phase is stabilized down to $U_c^I=4.72$.
Although this result is the same as that in the system 
with $R=1$,
it does not necessarily imply that 
the difference of the bandwidths is irrelevant in the whole parameter space.
Indeed, further increase of the Hund coupling suppresses orbital fluctuations,
leading to double transitions in the system.
When $J/U=0.1$, the introduction of the Coulomb interaction decreases 
the orbital and charge susceptibilities.
Consequently, these quantities merge to each other and 
the spin susceptibility diverges at the critical point 
$U_c^M=3.34$.
This means that the second-order OSMT occurs in the narrower band.
Thus, free spins $s=1/2$ are induced at each site, 
yielding the residual entropy $S/L=\ln 2$.
Furthermore, as the interactions are increased, 
both orbital and charge susceptibilities reach zero at $U_c^I=4.06$. 

\begin{figure}[htb]
\includegraphics[width=\linewidth]{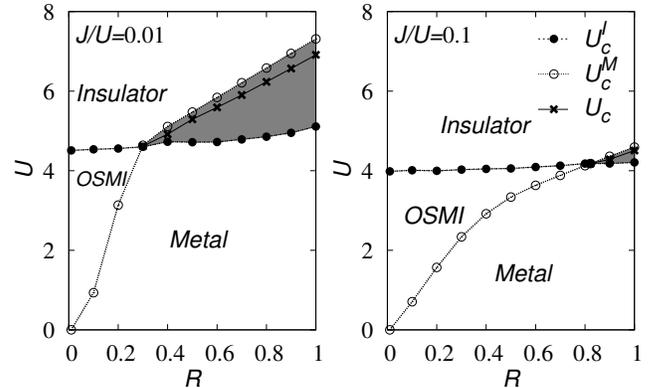}
\caption{The phase diagram when $J/U=0.01$ and $J/U=0.1$. 
The gray region denotes 
the coexisting state.
}\label{fig:R_U}
\end{figure}

Performing similar calculations with several choices of $R$,
we end up with the phase diagrams shown in Fig. \ref{fig:R_U}.
An important point is that 
the phase boundaries $U_c^M$ and $U_c^I$ 
cross each other at $R=R_c$,
where $R_c=0.30$ and $0.82$ for $J/U=0.1$ and $0.01$, respectively.
When $R<R_c$, the difference of the bare bandwidths is essential
to stabilize the OSMI phase between two phase boundaries $(U_c^M< U < U_c^I)$.
However, when $R>R_c$,
the metallic phase and the Mott insulating phase
coexist in the region $U_c^I<U<U_c^M$.
In this case, the nature of Mott transitions is similar to 
that in the system with same bandwidths, 
where the first-order Mott transition occurs.
The above result may suggest that the existence of the OSMT is associated with 
the first-order transition in the multi-orbital Hubbard model 
with same bandwidths.

\begin{figure}[htb]
\includegraphics[width=\linewidth]{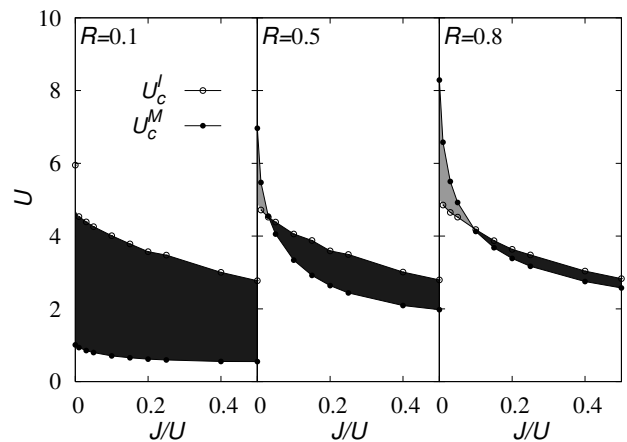}
\caption{The phase diagram for the degenerate Hubbard model 
with different bandwidths $R=0.1$, $0.5$ and $0.8$.
The black and gray region denote 
the OSMI phase and the coexisting state, respectively
}\label{fig:Up_U}
\end{figure}

To discuss 
how the coexisting phase competes with the OSMI phase,
we also show the phase diagrams with a fixed $R$ in Fig. \ref{fig:Up_U}.
When $R<0.192$, 
there always appears the OSMI phase 
due to the large difference of bandwidths. 
Note that the the phase boundary sometimes has a discontinuity at $J/U=0$,
which may originate from the spin and orbital degeneracy 
of the Mott insulating ground state, 
as discussed in the previous section.
In contrast, 
the OSMI phase and the coexisting region appear in the case $0.192<R<1$.
When the system has the large Hund coupling, 
orbital fluctuations are suppressed and
the OSMI phase is stabilized.
The decrease of $J$ enhances orbital fluctuations in the metallic phase, 
making the OSMI phase unstable.
The phase boundaries then merge at the critical point.
Further decrease of the Hund coupling induces 
a first-order single Mott transition due to enhanced orbital fluctuations,
where the coexisting phase appears instead of the OSMI phase.
When $R\rightarrow1$, the difference of the bandwidths becomes irrelevant.
Therefore, the region of the OSMI phase shrinks and 
the coexisting phase becomes stable.

Up to now, there were some controversial conclusions for the nature of the OSMT.
Most of studies claimed that the second order OSMT occurs in general.\cite{Koga04,Inaba05os}
In contrast,
Medici {\it et al.}\cite{Medici05} and Arita {\it et al.}\cite{Arita05} 
recently claimed that the OSMT is 
of first order by combining DMFT with the exact diagonalization and 
the projective quantum Monte Carlo simulations.
In our paper, we have calculated the grand potential in the system
carefully to clarify that the OSMT is of second order.
Although the accuracy of our calculations depends on 
the size of the reference system in the framework of the SFA, 
the nature of Mott transition is hardly affected by the size, as discussed before.
Therefore, we believe that
the obtained results may shed light on the nature of the Mott transitions in
the two-orbital systems.

\section{Summary}\label{sec:summary}
We have investigated the two-orbital Hubbard model in infinite dimensions
to discuss the nature of Mott transitions at half-filling.
By making use of the self-energy functional approach,
we have clarified that the introduction of the Hund coupling induces
the first-order transition in the system with same bandwidths.
It has also been found that the increase of the Hund coupling
suppresses orbital fluctuations, where the singularity characteristic of 
the first-order transition becomes obscure.
Namely, the upper and lower critical points for the transition 
approach each other closely, where the weak first-order transition occurs.
This behavior is also affected by the difference of the bandwidths.
In fact, the difference of the bandwidths suppresses orbital fluctuations, 
making the first-order transition unstable.
As a consequence, the double second-order transitions occur, 
where the OSMI phase
appears between the metallic and the Mott insulating phase.
Our obtained results reproduce some previous works 
on the system with same and different bandwidths.
\cite{Ono03,Pruschke05,Ferrero05,Medici05}
Furthermore, taking into account of orbital fluctuations carefully,
we have resolved some controversial conclusions 
for the nature of Mott transitions 
at half-filling.

\begin{acknowledgments}
We would like to thank S. Suga, N. Kawakami, M. Sigrist, T. M. Rice 
and A. Liebsch for valuable discussions.
Numerical computations were carried out at the Supercomputer Center, 
the Institute for Solid State Physics, University of Tokyo. 
This work was supported by a Grant-in-Aid for Scientific Research from 
the Ministry of Education, Culture, Sports, Science, and Technology, Japan.
\end{acknowledgments}


\end{document}